\newcommand{\text}{\rm}
\newcommand{\fg}[1]{Figure\,\ref{fig:#1}}

\newcommand{\lna}{\ln\!A}
\newcommand{\mlna}{\langle \lna \rangle}
\newcommand{\xmax}{X_{\rm max}}
\newcommand{\nmu}{{N_\mu}}

\documentclass{PoSa}

\usepackage{lineno}
\title{Working Group Report on the Combined Analysis of Muon Density Measurements from Eight Air Shower Experiments}

\ShortTitle{WHISP: Working group in Hadronic Interactions and Shower Physics}

\author{\speaker{Lorenzo Cazon},
  for the EAS-MSU$^1$, 
  IceCube$^2$, 
  KASCADE-Grande$^3$, 
  NEVOD-DECOR$^4$, 
  Pierre Auger$^5$, 
  SUGAR$^6$, 
  Telescope Array$^7$, 
  and Yakutsk EAS Array$^8$ 
  collaborations\footnote{for collaboration lists see %\pos
    {PoS(ICRC2019)1177}}\\
  $^*$ LIP, Av. Prof. Gama Pinto 2, Lisbon, Portugal   E-mail: \email{cazon@lip.pt}\\
  $^1${EAS-MSU experiment, Moscow, Russia},\\
  $^2${IceCube Neutrino Observatory, Madison, USA
    \href{https://icecube.wisc.edu/collaboration/authors/icrc19_icecube}{\rm https://icecube.wisc.edu/collaboration/authors/icrc19\_icecube}},\\
  $^3${KASCADE-Grande experiment, Karlsruhe, Germany 
    \href{https://web.ikp.kit.edu/KASCADE}{\rm https://web.ikp.kit.edu/KASCADE}},\\
  $^4${NEVOD-DECOR experiment, Moscow, Russia},\\
  $^5${Pierre Auger Observatory, Malarg\" ue, Argentina \\
   Full author list: Pierre Auger Collaboration and additional author(s): \href{http://www.auger.org/archive/authors_icrc_2019_a.html}{\rm http://www.auger.org/archive/authors\_icrc\_2019\_a.html}},\\
    $^6${SUGAR Array, Sidney, Australia},\\
  $^7${Telescope Array Project, Salt Lake City UT, USA 
    \href{http://www.telescopearray.org/research/collaborators}{\rm http://www.telescopearray.org/research/collaborators}},\\
  $^8${Yakutsk EAS Array, Yakutsk, Russia 
\href{https://ikfia.ysn.ru/en/theyakutskarrayteam}{\rm https://ikfia.ysn.ru/en/theyakutskarrayteam}}\\
}

\abstract{We present a meta-analysis of recent muon density measurements made by eight air shower experiments which cover shower energies ranging from PeV to tens of EeV regarding the muon puzzle in extensive air showers. Some experimental analyses reported deviations between recorded and simulated muon densities in extensive air showers, and others reported no discrepancies. Comparisons between experiments were made using a universal reference scale based on the relative difference to simulated proton and iron initiated air showers. We have applied a cross-calibration of energy scales between experiments based on the isotropic flux of cosmic rays as a reference. Above 10 PeV, most experimental data show a muon excess with respect to simulated air showers, including those performed with the recent post-LHC high-energy interaction models. The discrepancy increases with the shower energy with a slope 8 sigma away from the predictions by EPOS-LHC and QGSJet-II.04. The effect of measurements being made at different
  zenith angles and energy threshold of muons across different experiments will be addressed.}

\FullConference{36th International Cosmic Ray Conference -ICRC2019-\\
		July 24th - August 1st, 2019\\
		Madison, WI, U.S.A.}

\begin{document}
\setcounter{page}{2}
\section{Introduction}

The determination of the primary mass, arrival direction, and energy of cosmic rays arriving at Earth with energies exceeding $10^{15}\,{\rm eV}$ is done by means of detection and analysis of Extensive Air Showers (EAS). Two of the most used  observables to infer the mean logarithmic mass of the primary, $\mlna$, are the depth of the shower maximum in the atmosphere $\xmax$ and the number of muons in the shower, $\nmu$.  One of the dominant sources of systematics in $\mlna$ stems from the uncertainty in high energy hadronic interaction models used in air shower simulations. While $\xmax$ is linked to the electromagnetic component of EAS, $\nmu$ directly stems from the hadronic cascade, which is more prone to model uncertainties.   By checking the consistency of $\xmax$ and $\nmu$ in terms of mass interpretation, models can be tested, opening a window for particle physics in phase-space regions beyond the reach of the LHC. At the same time, models can be improved, allowing a more precise determination of the mass of the primaries.

In the year 2000, the HiRes/MIA collaboration reported a discrepancy of the number of muons in simulated and measured air showers between $10^{17}$ to $10^{18}$\,eV \cite{AbuZayyad:1999xa}. A decade later, NEVOD-DECOR reported \cite{Bogdanov:2010zz,Bogdanov:2018sfw} an increase of muon density relative to simulations, and also SUGAR array observed an excess in data~\cite{Bellido:2018toz}. EAS-MSU~\cite{Fomin:2016kul} reported no muon number discrepancy as well as KASCADE-Grande~\cite{Apel:2017thr}, which in addition reported differences in the muon number evolution with zenith angle with respect to the latest hadronic interaction models tuned to LHC-data. The Pierre Auger Observatory~\cite{Aab:2014pza,Aab:2016hkv} and Telescope Array~\cite{Abbasi:2018fkz} also observed a muon deficit in simulations, at $10^{19}$ eV. These measurements, preliminary data from \mbox{IceCube}~\cite{Gonzales:2018IceTop} and AMIGA~\cite{mueller2018}, and unpublished data from Yakutsk~\cite{yakutskpc} will be systematically compared. In this report, we present a meta-analysis of the muon density measurements performed by these eight EAS experiments.

\section{Measurement of the muonic density}

\begin{figure*}
\centering
\includegraphics[width=0.325\textwidth,clip,trim=45 10 10 0]{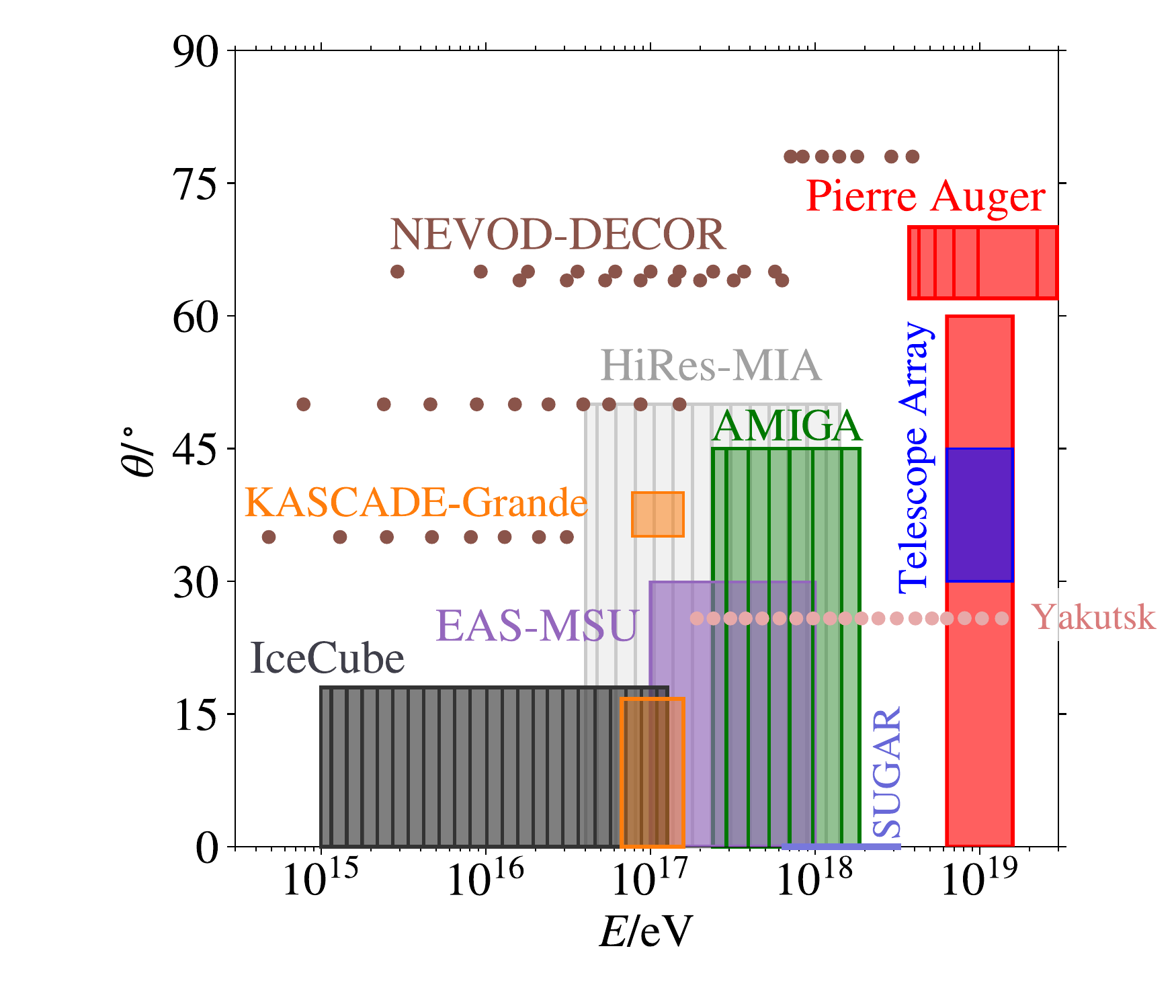}
\hfill
\includegraphics[width=0.31\textwidth,clip,trim=45 10 30 0]{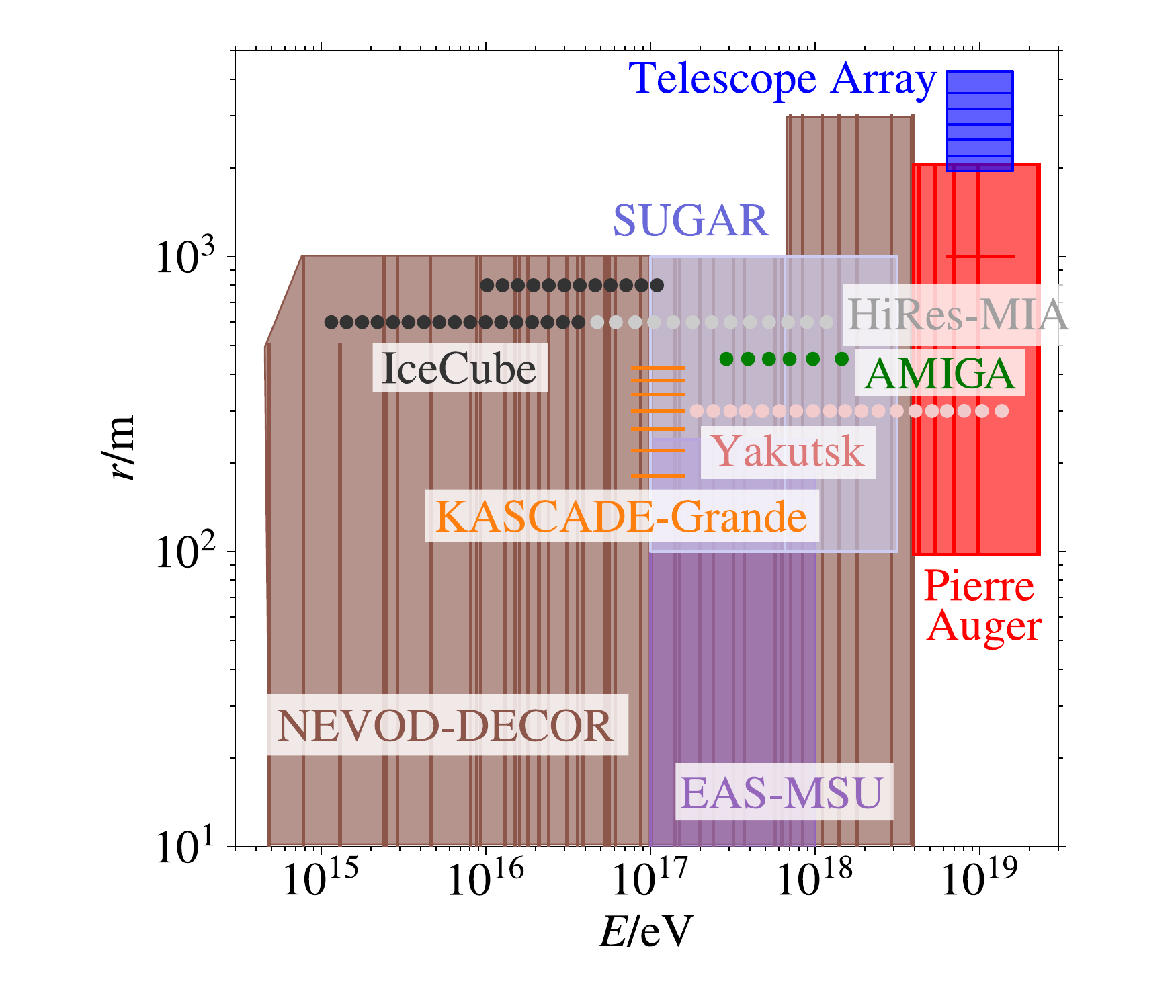}
\hfill
\includegraphics[width=0.32\textwidth,clip,trim=30 10 30 0]{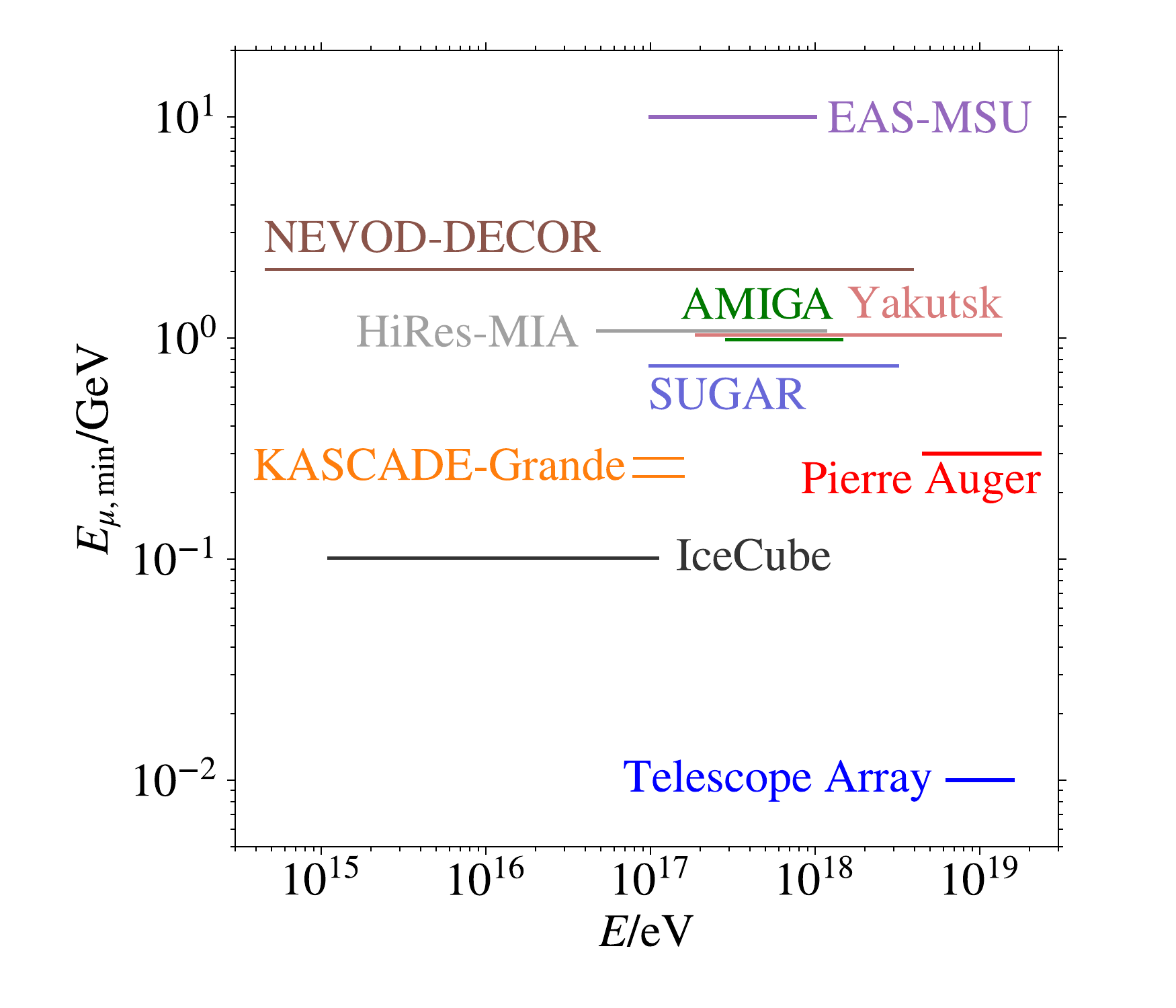}
\caption{Air shower experiments have measured the muon density at ground under various conditions, which are shown here as a function of the shower energy. Points and lines indicate a measurement in a narrow bin of the parameter, while boxes indicate integration over a parameter range. \emph{Left:} Zenith angle of air showers. \emph{Middle:} Lateral distance of the muon density measurement. \emph{Right:} Energy threshold for the muons that are counted in the experiment.
  }
\label{fig:phase_space}
\end{figure*}

The muon density at the ground depends on many parameters:
cosmic-ray energy $E$,
zenith angle $\theta$,
shower age (vertical depth $X$,  and zenith angle of the shower $\theta$),
lateral distance $r$ from shower axis, and 
energy threshold $E_{\mu,{\rm min}}$ of the detectors for muons.
The parameter space covered by each experiment is shown in \fg{phase_space}.
A direct comparison of the muon measurements is not possible, due to the very different conditions and techniques which were used.
Instead, we compare each experiment to air shower simulations in the same observation conditions by means of a data/MC ratio.

According to the type of shower energy and muon density reconstruction, we can classify the experiments into three groups:

\begin{description}
  \item[Shower energy and muon density] The Pierre Auger Observatory (including its muon detector AMIGA), Telescope Array, and the Yakutsk experiment are all capable of measuring the shower energy, $E_{\rm cal}$, using Cherenkov or fluorescence light, which can be converted into the primary energy, $E$, with little model dependence. These experiments come close to measuring the primary cosmic-ray energy, $E$, independently of the muon density at the ground.%, and can compute the data/MC ratio for showers with the same energy.

  The IceCube Neutrino Observatory can also be classified under this category, although it does not observe the air shower optically. It can measure the shower energy with low model dependence thanks the detector being close to the average depth of the shower maximum~\cite{Aartsen:2013wda}.

\item[Muon and electron density] KASCADE-Grande and EAS-MSU measure signals from electrons and muons separately. These experiments compute the data/MC ratio for showers in the same electron-density interval, which can contain different primary energies, compensated by the attenuation produced by differences in the depth of the shower maximum.

\item[Muon density only] NEVOD-DECOR and SUGAR are pure muon detectors, without a separate energy estimator.
    The flux of showers is measured in intervals of muon density, which is then compared with a simulated flux, computed from an external model.  NEVOD-DECOR uses an average cosmic-ray flux computed from multiple experiments, while SUGAR uses the flux measured by the Pierre Auger Observatory.
\end{description}

While most of the shower parameters can be easily matched in simulation and experiment, the cosmic-ray energy, $E$, is difficult to match.
The number of muons scales as $\nmu \propto A^{1-\beta} E^\beta$, with $\beta \simeq 0.9$~\cite{Matthews:2005sd}. This could cause two experiments with an energy-scale offset of 20\,\% to have a 18\,\% additional offset in the data/MC ratios.
If we take the cosmic ray flux as a universal reference, and assuming that all deviations in measured fluxes between different experiments arise from energy-scale offsets, a relative energy-scale ratio $E_{\rm data}/ E_{\rm ref}$ can be found for each experiment so that the all-particle fluxes overlap. 
This approach is well-known and has been applied successfully in other work~\cite{Hoerandel:2002yg,Dembinski:2017zsh}.

\begin{table}
\centering
\caption{Table showing the energy-scale adjustment factors obtained from cross-calibration (2nd column), the median $\sec \theta$ of several data sets (3rd column), and the minimum energy of the muons at production, due to energy loss in the atmosphere and detector shielding (see text for explanation).}
\label{tab:energy_scale_correction}       % Give a unique label
% For LaTeX tables you can use
\begin{tabular}{ll|lll}
%\hline
Experiment & $E_{\rm data} / E_{\rm ref}$ & $\sec \theta$ & $E_{\mu \, {\rm prod}}/$GeV \\
\hline
EAS-MSU & -                   & 1.1  & 11.9\\
IceCube Neutrino Observatory & 1.19 & 1.0 & 0.7\\
KASCADE-Grande & -            & 1.0 , 1.3 & 1.5 , \,2.1  \\

NEVOD-DECOR & 1.08                  & 2.3 , 4.8   & 8.4 , 18.6    \\
Pierre Auger Observatory  & 0.948 & 1.3 , 2.4 & 1.8 , \,4.0\\
AMIGA & 0.948 & 1.2 & 2.4 \\
SUGAR & 0.948 & 1.0 & 1.9 \\
Telescope Array & 1.052 & 1.3 & 1.4 \\
Yakutsk EAS Array & 1.24 & 1.1 & 2.6 \\
\hline
\end{tabular}
\end{table}

The cross-calibration factors used in this report are given in Table~\ref{tab:energy_scale_correction}, and are summarized in \cite{Dembinski:2019uta}.
We assume that the resulting overall energy-scale has an uncertainty of at least 10\,\%.

\subsection{Combined measurements}

\begin{figure*}
\centering
\includegraphics[width=\textwidth]{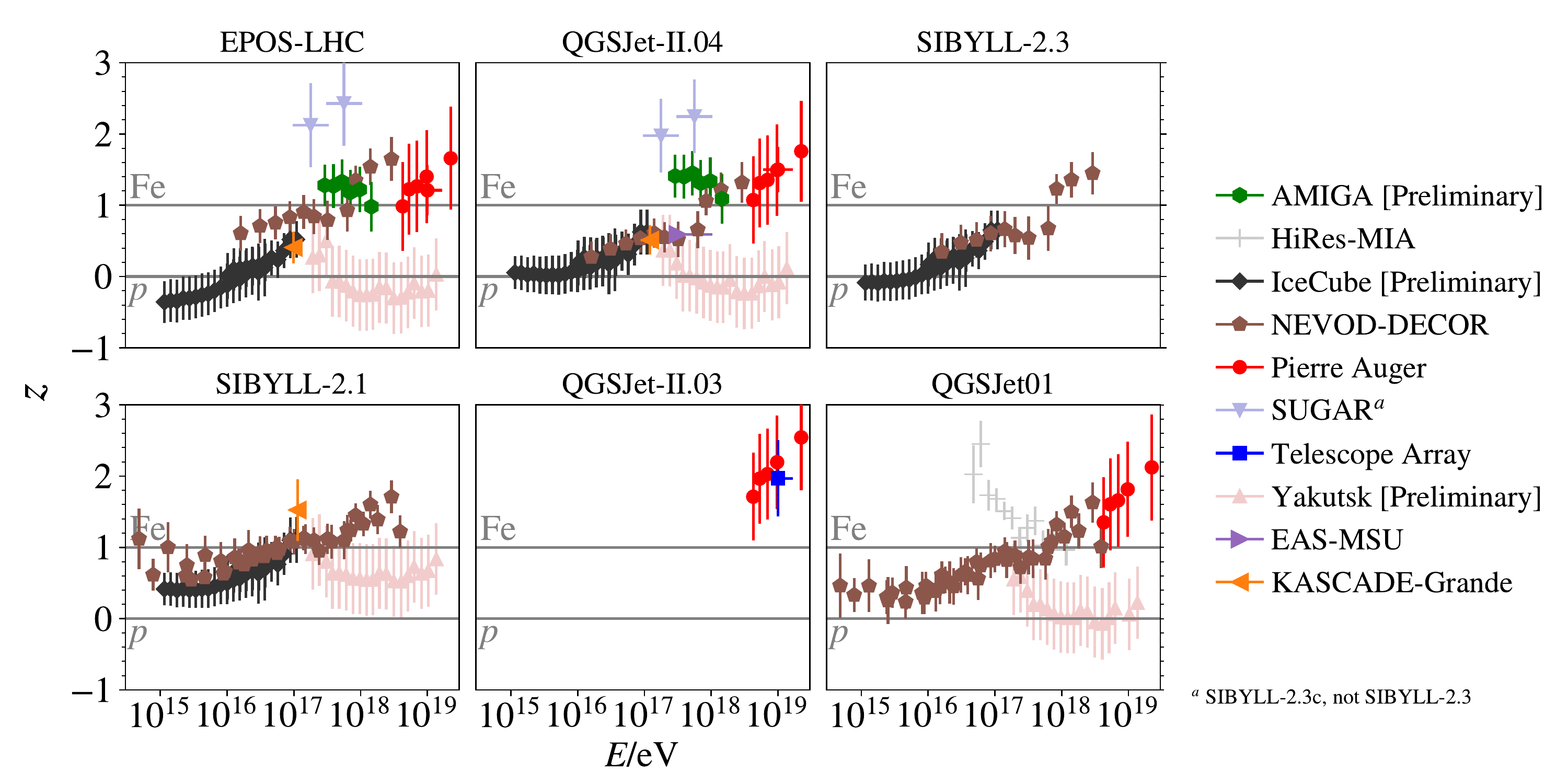}
\caption{Muon density measurements converted to the $z$-scale
  for each hadronic interaction model.
  When corresponding simulations are missing for an experiment, no points are shown. Error bars show statistical and systematic uncertainties added in quadrature (systematic uncertainties are dominant in nearly all measurements).}
\label{fig:muon_original}
\end{figure*}

\begin{figure*}
\centering
\includegraphics[width=\textwidth]{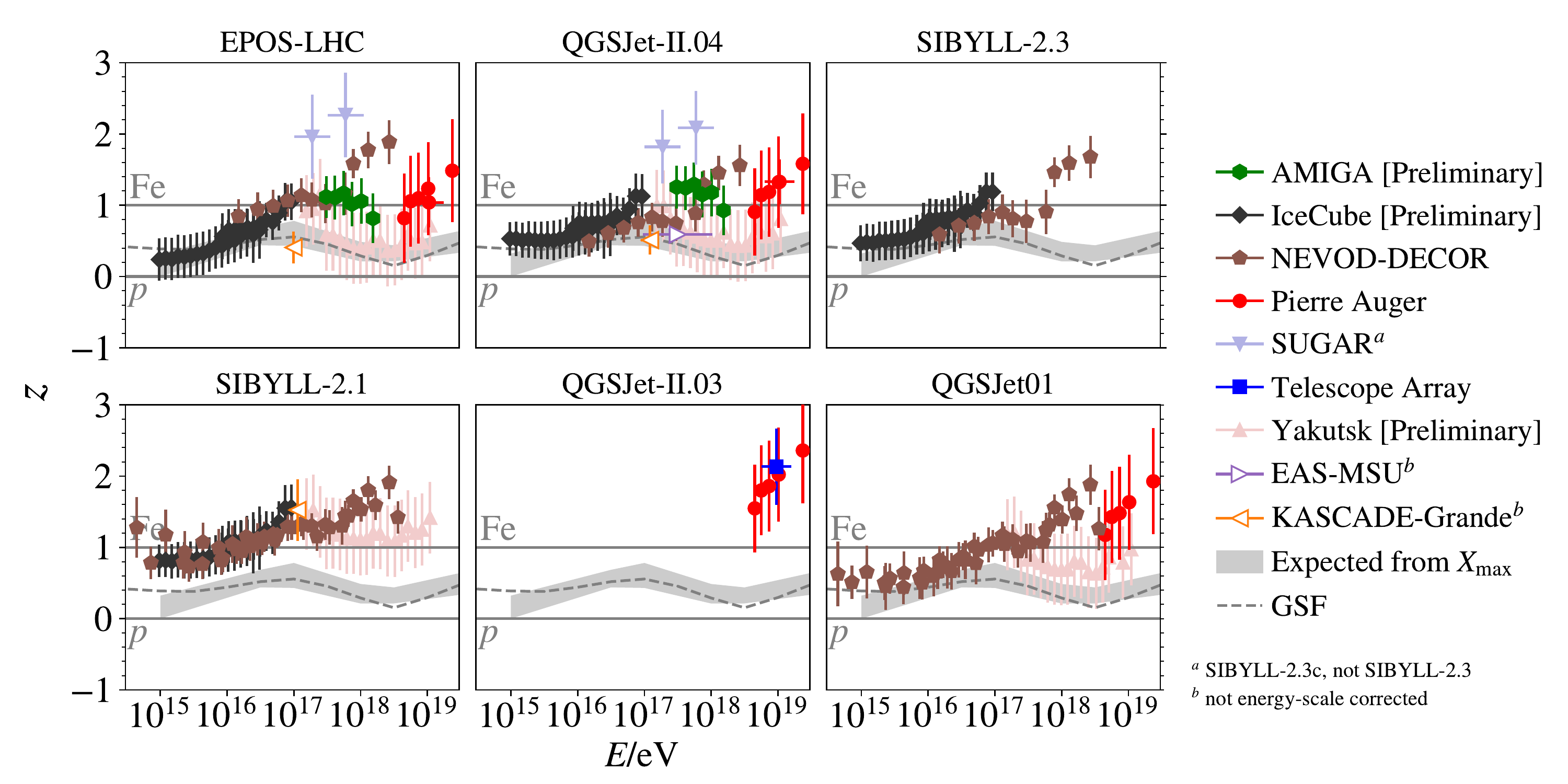}
\caption{Data from \fg{muon_original} after applying  the energy-scale cross-calibration. The points for KASCADE-Grande and EAS-MSU were not cross-calibrated in this report and are included for comparison. Shown for comparison are $z$-values expected for a mixed composition from optical measurements (band)
  based on ~\cite{Kampert:2012mx},
  and from the GSF model (dashed line) \cite{Dembinski:2017zsh}.}
\label{fig:muon_energy_rescaled}
\end{figure*}

To compare all the measurements, we introduce the $z$-scale
\begin{equation}
z = \frac{ \ln(\nmu^{\rm det}) - \ln(\nmu_p^{\rm det})}{ \ln(\nmu_{\rm Fe}^{\rm det}) -  \ln(\nmu_p^{\rm det})},
\label{eq:z}
\end{equation}
where $\nmu^{\rm det}$ is the muon density estimate as seen in the detector, while $\nmu_p^{\rm det}$ and $\nmu_{\rm Fe}^{\rm det}$ are the simulated muon density estimates for proton and iron showers, accounting for detector effects. The $z$-scale ranges from 0 (pure proton showers) to 1 (pure iron showers),  if there is no discrepancy between real and simulated air showers, in which case one could use  $\mlna =z \, \ln 56$ to infer the average logarithmic mass.

Shown in \fg{muon_original} are the converted measurements for each hadronic interaction model. Note that the conversion to $z$ is only possible when $\nmu_{\rm p}^{\rm det}$ and $\nmu_{\rm Fe}^{\rm det}$ are available for that model under the specific experimental conditions. The scatter is drastically reduced after the cross-calibration, as can be seen in \fg{muon_energy_rescaled}. The points  move horizontally by the relative amount $(E_{\rm data}/E_{\rm ref})^{-1}$ and vertically by $z_{\rm ref}-z_{\rm data}=\frac{\beta\ln (E_{\rm data}/E_{\rm ref})}{\ln(\nmu_{\rm Fe}^{\rm det}) - \ln(\nmu_p^{\rm det})}$.
As expected, the cross-calibration improves the agreement between data from different experiments. The points for KASCADE-Grande\footnote{Latest KASCADE-Grande results, presented in this conference \cite{KASCADEICRC19}, were not yet included in this comparison.}
 and EAS-MSU were not cross-calibrated in this report and are included for comparison.
We emphasize again that the reference energy-scale after cross-calibration has a remaining uncertainty of at least 10\,\%. This means that $z$-values in all plots can be collectively varied by about $\pm 0.25$.

Now, we consider the effect of an energy-dependent mass composition.
Shown in \fg{muon_energy_rescaled} is a band which contains the optical measurements of the depth $\xmax$ of shower maximum from several experiments after conversion into $\mlna$ based on air shower simulations with EPOS-LHC. We will use this as an  estimate of the mass composition. The band is independent of the muon measurements and can be used as a reference. Also shown is the $z_{\rm mass}=\mlna/\ln56$ value computed from the GSF model \cite{Dembinski:2017zsh}, which is based on optical and muon measurements and averages over experiments and model interpretations of air shower data.

It can be seen that the measured $z$ values do not follow $z_{\rm mass}$, implying that the models do not consistently describe the muon density.
EPOS-LHC \cite{Pierog:2013ria}, QGSJet-II.04 \cite{Ostapchenko:2013pia}, and SIBYLL-2.3 \cite{Riehn:2017mfm},
and the pre-LHC model QGSJet01
give a reasonable description of data up to a few $10^{16}$\,eV. At higher shower energies, a muon deficit in simulations is observed ($z > z_{\rm mass}$) in all models.
Shown in \fg{zfit} is the difference $\Delta z = z - z_{\rm mass}$ for EPOS-LHC and QGSJet-II.04, the two latest-generation models with most cross-calibrated data points in this study. Subtracting $z_{\rm mass}$ is expected to remove the effect of the changing mass composition. An energy-dependent trend in $\Delta z$ remains.

\subsection{Energy-dependent trend}

\begin{figure*}
\centering
\includegraphics[width=0.49\textwidth,clip,trim=30 10 30 10]{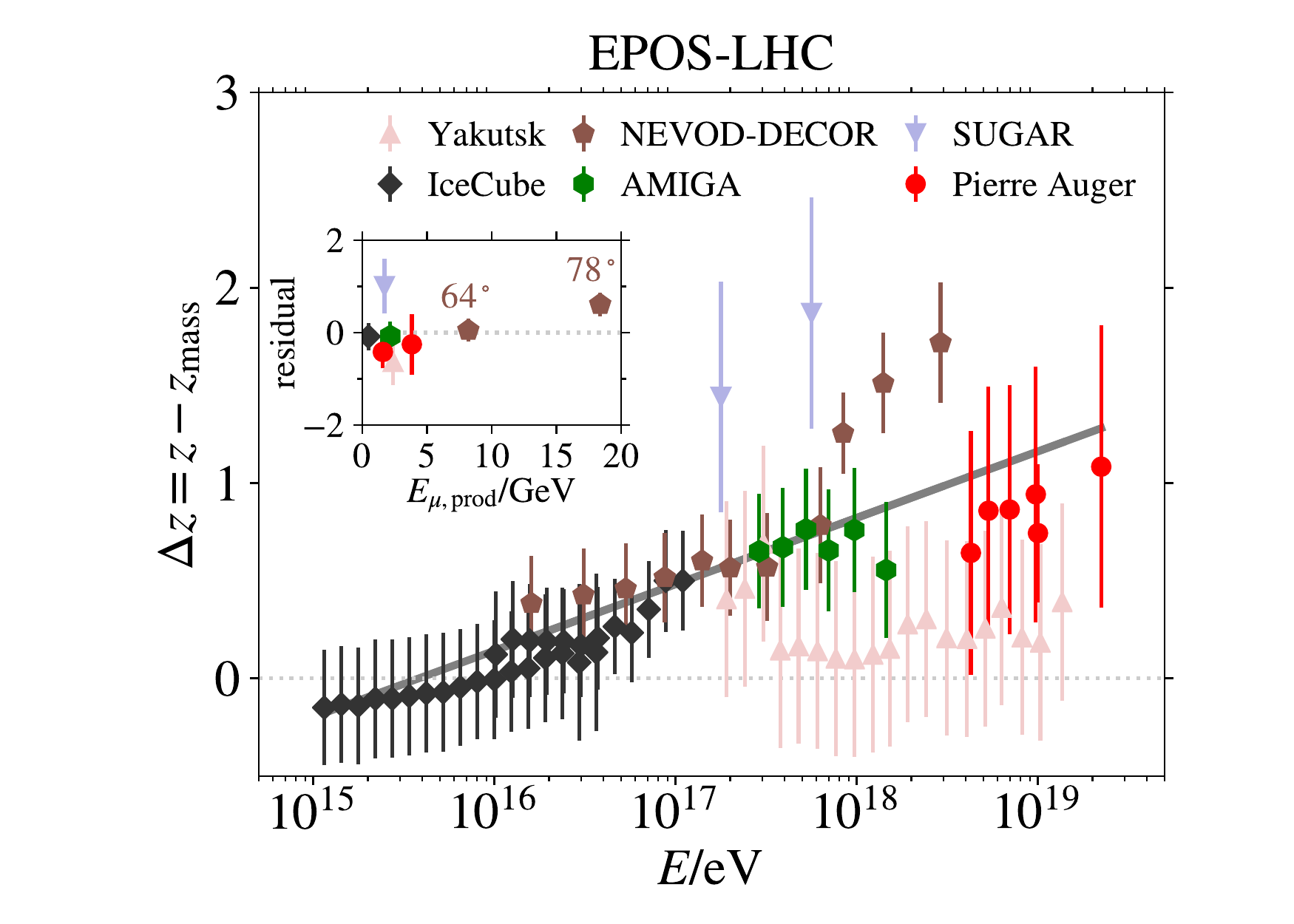}
\hfill
\includegraphics[width=0.49\textwidth,clip,trim=30 10 30 10]{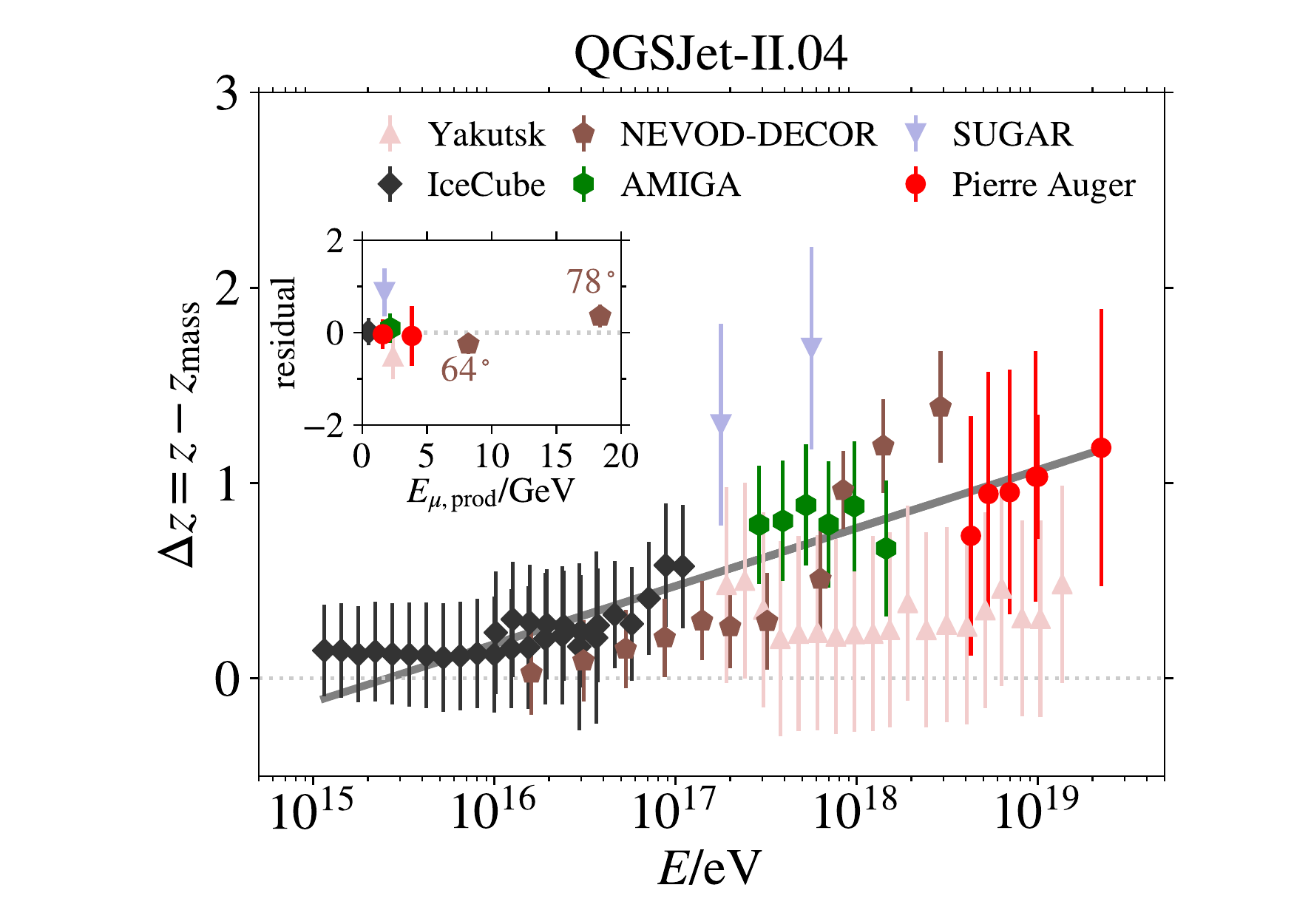}
\caption{$\Delta z=z-z_{\rm mass}$ for EPOS-LHC and QSGJet-II.04.
  The function $\Delta z_{\rm fit} = a + b \log_{10}(E/10^{16}{\rm eV})$ was fitted assuming a correlation coefficient $\alpha=0.5$ within each data-set, resulting in $b=0.34 \pm 0.04$ for EPOS-LHC and $b=0.30 \pm 0.03$ for QGSJetII.04. The inset shows the average residual per data set with respect to the fitted line, $\Delta z-\Delta z_{\rm fit}$,  as a function of the minimum energy of muons at the production point high in the atmosphere, $E_{\mu \, {\rm prod}}$ . (See text for explanations).}
\label{fig:zfit}
\end{figure*}

To quantify the observed trend in $\Delta z$ as a function of energy, the function
\begin{equation}
\Delta z_{\rm fit} = a + b \, \log_{10}(E/{10^{16}\rm eV}),
\label{eq:fit}
\end{equation}
is fitted to the data shown in \fg{zfit}, with free parameters $a$ and $b$.
The uncertainty of $b$ scales with the uncertainties of the data. The error bars of most data points are dominated by systematic uncertainties, which are correlated for data points from a single set.
The exact amount of correlation is not known. We work around this problem by repeatedly fitting the data under different correlation assumptions.
We use the least-squares method assuming
a correlation factor $\alpha$ between points belonging to the same data set and 0 otherwise.
The fit was repeated for values of $\alpha$ from 0 to 0.95.
To adjust for over- or underestimated uncertainties in the input, the raw result was rescaled with the $\chi^2$ value and the degrees of freedom $n_{\rm dof}$ of the fit, $\sigma_b = \sigma_b^{\rm raw} \sqrt{\chi^2/n_{\rm dof}}$~\cite{James:2006zz}. For EPOS-LHC the slope ranges from $b=0.29\pm0.04$ up to $b=0.35\pm0.04$ for $\alpha=0$ and $\alpha=0.95$ respectively. For QGSJet-II.04, it ranges from $b=0.22\pm0.03$ up to $b=0.31\pm0.04$ for $\alpha=0$ and $\alpha=0.95$ respectively. In both cases, the significance is always larger than 8-sigma.

As previously mentioned, each experiment has a shielding material $\Delta X_{\rm det}$ above the actual detector volume, which effectively translates into an energy threshold for muons $E_{\mu {\rm min}}(\theta)=\frac{dE_\mu}{dX} \Delta X_{\rm det} \sec(\theta)=E_{\mu {\rm min}} \sec(\theta)$ due to energy loss $\frac{dE_\mu}{dX} \simeq 2$ MeV / g cm$^{-2}$, above the detector. Energy losses in the atmophere are also taken into account. Most of the muons are produced approximately at the maximum of the muon production depth distribution, around\footnote{In \cite{Aab:2014dua}, the observed value is $\sim 475$ g cm$^{-2}$ above $2 \times 10^{19}$ eV. The elongation rate for proton and iron  simulations is around 25 g cm$^{-2}$/decade, which then sets a lower bound of $\sim 300$ g cm$^{-2}$ at $10^{15}$ eV. It must be also noted that the {\it apparent} muon production depth (MPD) depends on the particular condition of observations, contrary to the {\it total/true} MPD \cite{Cazon:2012ti}. Therefore, we have chosen the central value $400$ g cm$^{-2})$. Assuming an uncertainty of $\sim 100$ g cm$^{-2}$, the error due to these approximations is around $\pm$0.2 GeV in $E_{\mu \, {\rm prod}}$.} $\sim 400$ g cm$^{-2}$. Therefore $E_{\mu \rm atm}(\theta)=\frac{dE_\mu}{dX} (X \sec(\theta)-400$ g cm$^{-2})$, where $X$ is the vertical atmospheric depth of the considered experiment. The minimum energy required at production for a muon to be detected in each experiment is then $E_{\mu \, {\rm prod}}= E_{\mu {\rm min}}(\theta)+E_{\mu \rm atm}(\theta)$. Table~\ref{tab:energy_scale_correction} shows this energy for the different data sets. When we scan for different $E_{\mu \, {\rm prod}}$ we are in fact looking for differences in the energy spectrum of muons at production \cite{Espadanal:2016jse}. Inset of \fg{zfit} displays the residuals $\Delta z-\Delta z_{\rm fit}$  as a function of $E_{\mu \, {\rm prod}}$.

We have modified equation \ref{eq:fit} and added an extra parameter to account for this possible effect, as $\Delta z_{\rm fit} = a + b\log_{10}(E/{10^{16} \rm eV})+c \, E_{\mu \, {\rm prod}}$.  The results of this modification are not conclusive because  $c$ changes significantly when the data set at $\sec(78^\circ)=4.8$ from NEVOD-DECOR is excluded from the fit. 
At these extreme zenith angles, the average density of the atmosphere drops very rapidly in the region where the hadronic cascade develops. Therefore, the critical energy of the $K/\pi$ mix of the shower will change. A possible mismatch with respect to models might also produce visible effects, and it is difficult to disentangle the effects in the current state of analysis. Another complication is the effective  distance of the muon bundles recorded by NEVOD-DECOR to the shower core and its evolution with primary energy and zenith angle. This must be carefully accounted for, as it might introduce a further mismatch with respect to simulations.

\section{Summary and outlook}

A comprehensive collection of muon measurements in air showers with energies from PeV up to tens of EeV is presented. We developed the $z$-scale as a comparable measure of the muon density between different experiments and analyses. The $z$-scale uses air shower simulations as a reference to compare muon density measurements taken under different conditions. We cross-calibrated energy-scales of experiments when possible, using the isotropic all-particle flux of cosmic rays as a reference.

A remarkably consistent picture was obtained. Muon measurements seem to be consistent with simulations based on the latest generation of hadronic interaction models, EPOS-LHC, QGSJet-II.04, and SIBYLL-2.3, up to about $10^{16}$\,eV. At higher energies, a growing muon deficit in the simulations is observed. The slope of this increase in $z$ per decade in energy is 0.22 to 0.35 for EPOS-LHC and QGSJet-II.04, with 8\,$\sigma$ significance.

We have looked for other trends in the deviations between simulations and data.  We have calculated the approximated minimum energy of muons at production for each experiment to be detected. This value ranges from $\sim 1$ GeV up to $\sim 19$ GeV. Current data do not provide conclusive evidence of a deviation of the muon spectrum at production with respect to models. In this conference \cite{KASCADEICRC19}, KASCADE-Grande supports previous findings about a problem with the evolution of the muon content with the zenith angle between 10 PeV and 1 EeV.

The new generation of ongoing experimental upgrades and more refined analyses are expected to further reduce the uncertainities of the different experimental results. 

\bibliography{refs}
%\begin{thebibliography}{99}
%\bibitem{...}
%....
%\end{thebibliography}

\end{document}